\documentclass[twocolumn,aps,prd,amsmath,amssymb]{revtex4}
\bibpunct{}{}{,}{s}{}{,}
 
 \usepackage{amssymb,amsfonts}
\usepackage{graphicx,psfrag}

\newcommand{\be}{\begin{equation}}
\newcommand{\ee}{\end{equation}}
\newcommand{\bea}{\begin{eqnarray}}
\newcommand{\eea}{\end{eqnarray}}
\newcommand{\bdm}{\begin{displaymath}}
\newcommand{\edm}{\end{displaymath}}
\newcommand{\lb}{\label}

\newcommand{\I}{\mbox{i}}
\newcommand{\D}{\mbox{d}}

\begin{document}

\title{Why do cosmological perturbations look classical to us?}

\author{Claus Kiefer}
\email{kiefer@thp.uni-koeln.de}
\affiliation{Institut f\"ur Theoretische Physik, Universit\"at zu K\"oln,
 Z\"ulpicher Strasse~77, 50937 K\"oln, Germany}
\author{David Polarski}
\email{polarski@lpta.univ-montp2.fr}
\affiliation{Laboratoire de Physique Th\'eorique et Astroparticules, CNRS, 
 Universit\'e de Montpellier II, 34095 Montpellier, France}

\begin{abstract}
According to the inflationary scenario of cosmology, all structure in
the Universe can be traced back to primordial fluctuations during
an accelerated (inflationary) phase of the very early Universe.
A conceptual problem arises due to the fact that the
primordial fluctuations are quantum, while the standard scenario of
structure formation deals with classical fluctuations. In this essay
we present a concise summary of the physics describing the
quantum-to-classical transition. 
We first discuss the observational indistinguishability between classical 
and quantum correlation functions in the closed system approach (pragmatic 
view). We then present the open system approach with environment-induced 
decoherence. We finally discuss the question of the fluctuations' entropy 
for which, in principle, the concrete mechanism leading to decoherence 
possesses observational relevance.
\bigskip

\end{abstract}

\maketitle

%%%%%%%%%%%%%%%%%%%%%%%%%%%%%%%%%%%%%%%%%%%%%%%%%%%%%%%%%%%%%%%%%%%%
\section{Introduction}

It is often emphasized these days that the field of 
cosmology has entered a golden age. 
There is no doubt that the main reason for this statement is the 
accumulation of observations of ever increasing accuracy. 
In this way cosmological models aiming to describe the 
evolution of the Universe from the Big Bang until today 
are no longer purely speculative: their predictions can 
be tested and some models can indeed be ruled out. 

With the advent of inflationary models, according to which the Universe
underwent a phase of accelerated expansion at a very early stage, we
now have at our  
disposal theoretical tools to apprehend such fundamental 
problems as the origin of cosmological perturbations and 
the eventual formation of large-scale structures like 
galaxies. There are many ways in which inflationary models 
address fundamental physical theories. As
inflation is supposed to take place at very high energies 
in the early Universe, these models offer a unique window 
on energy scales of the order of $10^{15}$ GeV. Another 
intriguing aspect of these models is that inflationary 
perturbations originate from quantum fluctuations though 
we do not see this quantum nature in the Universe nowadays. 
It is this aspect of inflationary perturbations that we want 
to describe in our essay. 

We could, of course, as well consider non-inflationary 
cosmological models in which perturbations are assumed to be classical 
from the beginning on. However, such models are plagued with 
problems of causality as distant points on the last-scattering 
surface, about $350.000$ years after the Big Bang, were never 
in contact before. Hence the impressive homogeneity of the Cosmic 
Microwave Background (CMB) would have to be put in by hand in the absence 
of an inflationary stage. Inflationary
models are thus much more natural -- and they can be observationally
tested. 

The main part of our essay consists of four parts. We shall first give
in Sec.~II a brief review of inflationary cosmology and its
mechanism for the generation of perturbations. 
We then discuss in Sec.~III the quantum-to-classical transition 
in the closed system approach (we call it also the pragmatic view)
which focusses on the indistinguishability of quantum expectation 
values and classical stochastic averages. 
Sec.~IV presents the successful
observational predictions which emerge from this scenario. Sec.~V,
then, is devoted to environmental decoherence. We discuss the problem
of the classical variables (the pointer basis) as well as the 
entropy of the fluctuations and its observational significance.
We end with a brief conclusion.

%%%%%%%%%%%%%%%%%%%%%%%%%%%%%%%%%%%%%%%%%%%%%%%%%%%%%%%%%%%%%%%%%%%%%%%%% 
\section{Inflation}
We give here a brief review of the way in which inflationary models
give an elegant  
solution to many fundamental problems occuring in non-inflationary 
Big-Bang cosmology, see, for example, \cite{LL}. 
As we shall see, these models do also make 
characteristic predictions, by which we mean that in the absence of 
certain observable signatures most if not all inflationary models 
would be ruled out. We shall first describe the evolution of the
homogeneous background for inflation and then turn to the generation
of perturbations.

\subsection{Background expansion}

The crucial point here is that inflation is a stage of accelerated 
expansion. In this stage, proper (physical) scales are stretched by 
a huge factor so that scales inside the Hubble radius during inflation 
will eventually end up at the end of inflation far outside the Hubble 
radius. Today these scales can correspond to cosmological scales, 
and typically scales corresponding to the Hubble radius today have exited 
the Hubble radius during inflation about $65$ e-folds before the end 
of inflation. Typically, inflationary stages are quasi-de~Sitter
stages during which the Hubble parameter is nearly constant. As we shall 
see below, inflation provides a mechanism for the causal generation of
perturbations.

It is a basic assumption that our Universe is on large scales
homogeneous and isotropic. The metric is of the form
\be
ds^2 = dt^2 - a^2(t)~\left[ \frac{dr^2}{1-kr^2} + r^2 (d\theta^2 +
  \sin^2\theta~d\phi^2)\right]~;
\ee
a spatially flat universe corresponds to $k=0$, a closed universe to
$k=1$, and an open  
universe to $k=-1$. (We set the speed of light $c=1$ throughout.)
In an expanding universe, the scale factor $a(t)$ is a 
growing function of time, which starts close to zero at the Big Bang about 13.7 
billions years ago. The dynamics of the scale factor is given by the Friedmann 
equations,  
\bea
\biggl(\frac{\dot a}{a}\biggr)^2 &=& \sum_{i} \frac{8\pi G}{3} ~\rho_i
- \frac{k}{a^2}~,\lb{FR1}\\ 
\frac{\ddot a}{a} &=& -\frac{4\pi G}{3} \sum_{i} (\rho_i + 3 p_i)~,\lb{FR2}
\eea
where the index $i$ stands for any isotropic (comoving) perfect fluid. For 
radiation we have $p_{\rm r} = \rho_{\rm r}/3$, for dust $p_{\rm m} =
0$. For the recent  
accelerated expansion caused by some smooth dark energy component we would have 
$p_{\rm DE} = w_{\rm DE} ~\rho_{\rm DE}$, where $w_{\rm DE}< - 1/3$ is
still unknown and in many models time-dependent. 
>From (\ref{FR2}) the expansion is typically decelerated, ${\ddot a}<0$, unless 
at least one of the components satifies $\rho_i + 3 p_i<0$.

A space-independent scalar field $\phi(t)$ can be viewed as a comoving perfect 
fluid with  
\bea
\rho_{\phi} &=& \frac{1}{2} {\dot \phi}^2 + V(\phi)\lb{rho_phi}\ ,\\ 
p_{\phi} &=&  \frac{1}{2} {\dot \phi}^2 - V(\phi)~.\lb{p_phi}
\eea
Hence, a scalar field $\phi(t)$ can induce an accelerated expansion provided 
\be
{\dot \phi}^2 < V(\phi)~.\lb{acc} 
\ee
The field $\phi(t)$ driving the inflationary stage is called the inflaton 
and evolves according to the Klein-Gordon equation
\be
{\ddot \phi} + 3 H {\dot \phi} + \frac{dV}{d\phi} = 0~,\lb{KG}
\ee
which is the form taken by the conservation of energy for a perfect fluid 
defined by \eqref{rho_phi} and \eqref{p_phi}, and we have introduced the Hubble 
parameter $H\equiv {\dot a}/{a}$. 
In most inflationary models, the inflaton field 
$\phi(t)$ satisfies the slow-roll conditions ${\ddot \phi}\ll3 H {\dot
  \phi}$, and hence
\be
 3 H {\dot \phi} \approx - \frac{dV}{d\phi}~.
\ee 
It is easy to show that the conditions for slow-roll to hold are 
\be
\lb{slowroll}
{\dot H}\ll 3 H^2\ , \quad \frac{d^2 V}{d\phi^2}\ll 9 H^2~, 
\ee
in which case the condition (\ref{acc}) is amply satisfied so that 
accelerated expansion -- inflation -- takes place. 

We conclude this brief summary on the background evolution during 
inflation by discussing the relative evolution of physical scales. 
The Hubble radius $R_{\rm H}\equiv H^{-1}$ defines an important scale 
in cosmology. If $a\propto t^p$, we have $R_{\rm H}\propto t$, and it is clear 
that $R_{\rm H}$ grows faster than a physical scale $\lambda\propto a$ during 
a decelerated expansion, which has $p<1$. Hence physical scales greater 
than the Hubble radius, which we shall call ``superhorizon'' 
or ``super-Hubble'' scales, will eventually 
enter the Hubble radius, by which we mean that they will become 
smaller than $R_{\rm H}$: this is the situation in standard cosmology. 
This picture changes dramatically during inflation; to illustrate 
this we take a purely de~Sitter stage, which is characterized by 
$H$ = constant and 
$a(t)\propto \exp(Ht)$. Now it is clear that physical scales inside the 
Hubble radius, which we shall call ``subhorizon'' 
or ``sub-Hubble'' scales will eventually become larger 
than the Hubble radius.  

If a scale is said to cross the ``horizon'' 65 e-folds before the end 
of inflation, this means that 
at the end of inflation (where $t=t_{\rm e}$) one has 
$a=a_{\rm e} = {\rm e}^{65} a_k$ or $N_k=65$ with 
\be
N_k = \frac{a_{\rm e}}{a_k}\ ;
\ee
here, $a_k \equiv a(t_k)$ if $t_k$ is the ``horizon-crossing'' 
time of that particular scale with physical wavelength $(2\pi/k)a$. (Sometimes
the factor $2\pi$ is omitted.)
In a pure de~Sitter stage this would mean that $H(t_{\rm e} -
t_k)=65$. If we can  
compute the present physical scale evolving from the Hubble radius during 
inflation, we know to which physical scale today a scale with 
given $N_k$ corresponds. 
Depending on the details of the model, the Hubble radius 
{\em today} corresponds typically to a scale with $N_k\approx 65$. 
It can be shown that in slow-roll models $N_k$ can be computed from 
the value $\phi(t_k)$ and that it depends on the potential $V(\phi)$.

In consistent inflationary scenarios, inflation is 
followed by a standard cosmic expansion during which scales that 
went outside $R_{\rm H}$ become again smaller than $R_{\rm H}$; they 
``re-enter the horizon''. For a given scale, the number of e-folds 
between the first horizon crossing time $t_k$ during inflation 
and the second horizon crossing time
 during the radiation or matter stage at $t=t_{k,{\rm f}}$ is given by 
the parameter $r_k$, 
\be
\lb{rk}
r_k \equiv \ln \frac{a(t_{k,{\rm f}})}{a_k} \equiv \ln \frac{a_{k,{\rm
      f}}}{a_k}~. 
\ee 
We shall see in Sec.~III that $r_k$ coincides with the squeezing
parameter for a quantum state \cite{GS89}. 
For typical cosmological scales today, $r_k\sim 100$ and even larger.
Physically this corresponds to an enormous expansion 
of the universe, while a given scale $k$ was outside the Hubble radius. 
As we shall see below, the ensuing huge amount of squeezing 
for the quantum state plays a crucial role 
in the quantum-to-classical transition of inflationary quantum fluctuations. 
It also means that the quantum state originating from inflation is a very 
peculiar one. 
    
\subsection{Generation of perturbations}

During an inflationary stage, quantum field fluctuations evolve
according to the general principles of quantum field theory. Inflation 
is supposed to take place at an energy scale where space-time can be 
described as a classical curved space-time on which the quantum field 
fluctuations are defined. The inflaton fluctuations $\delta\phi({\mathbf x},t)$ 
can be treated as a massless scalar field. This is an excellent 
approximation when the inflaton field satisfies the slow-roll
conditions \eqref{slowroll}
and it is even exact when we consider primordial gravitational waves.

It is convenient to consider the rescaled quantity 
$a\delta\phi\equiv y({\mathbf x},t)$
and to work with conformal time $\eta = \int {dt}/{a(t)}$; a prime 
will be used to denote a derivative with respect to $\eta$. The
formalism presented here is exact for gravitational waves, but 
can be extended in a straightforward way to the primordial density
perturbations. 

The quantization of the real perturbation $y({\mathbf x},\eta)$ proceeds with 
the usual canonical quantization scheme. We start from the classical 
Hamiltonian describing the perturbations, 
\bea
H &\equiv& \int d^3{\bf x}~{\cal H}(y, p, \partial_i y, \eta)\nonumber\\
&=& \frac{1}{2}\int d^3{\bf k} \lbrack p({\bf k})p^*({\bf k})+k^2
y({\bf k})y^*({\bf k})\\ 
&+& \frac{a'}{a} \left(y({\bf k})p^*({\bf k})
+p({\bf k})y^*({\bf k})\right) \rbrack~,
\lb{Hcl}
\eea
where $p$ is the momentum conjugate to $y$,
\be
p\equiv \frac{\partial {\cal L}(y,y')}{\partial y'}  = y'-\frac{a'}{a}y~.
\ee
In (\ref{Hcl}) we have introduced the (time-dependent) Fourier transform
$y({\bf k},\eta)$ of the rescaled fluctuation $y({\mathbf
  x},\eta)$. (We sometimes keep the dependence on $\eta$.) In the Lagrangian 
formulation, it obeys the following classical equation of motion: 
\be
y''({\bf k},\eta)+\left(k^2- \frac{a''}{a} \right)y({\bf k},\eta)=0~.  \lb{eq}
\ee
Upon quantization, the Fourier transforms are promoted 
to operators on which we impose the canonical commutation relations,
\be
\lbrack y({\bf k},\eta),p^{\dag}({\bf k}',\eta) \rbrack =
{\rm i}\delta^{(3)} ({\bf k}-{\bf k}')~.\lb{ccy} 
\ee
(We set $\hbar=1$.)
We can write the Hamiltonian operator in the following way:
\bea
H & = & \int \frac{d^3{\bf k}}{2}  \bigl\lbrack k \bigl( a({\bf
  k})a^{\dag}({\bf k}) +  
                 a^{\dag}(-{\bf k})a(-{\bf k}) \bigr ) + \nonumber\\ 
  & &\; {\rm i} \frac{a'}{a} \bigl (a^{\dag}({\bf k})a^{\dag}(-{\bf k})-
                 a({\bf k})a(-{\bf k}) \bigr )\bigr\rbrack~.\label{H}~
\eea
The time-dependent annihilation operators $a({\bf k})$ 
(we often skip the argument $\eta$ for conciseness) appearing in 
(\ref{H}) are defined as usual,
\be 
a({\bf k})= \frac{1}{\sqrt{2}}
\left(\sqrt{k}~y({\bf k}) + \frac{\rm i}{\sqrt{k}}  p({\bf k})\right)~,
\ee
so that
\bea 
y({\bf k})&=&\frac{a({\bf k}) + a^{\dag}(-{\bf k})}{\sqrt{2k}}\ , \\
p({\bf k})&=&-{\rm i}\sqrt{\frac{k}{2}}\left(a({\bf k})-a^{\dag}({-\bf
    k})\right)~.\lb{yp} 
\eea
It is easily seen from (\ref{ccy}) that $a$ and $a^{\dagger}$
satisfy the commutation relations 
\be
\lbrack a({\bf k},\eta),a^{\dag}({\bf k}',\eta)\rbrack = \delta^{(3)}
({\bf k}-{\bf k}')~.\label{cca} 
\ee

Let us consider the time evolution of these operators. 
>From the Hamiltonian (\ref{H}) we get 
\be
\left(
\begin{array}{c}
a'({\bf k})\\
(a^{\dag}(-{\bf k}))'\\
\end{array}
\right)=
k~\left(
\begin{array}{cc}
-{\rm i}         &\frac{aH}{k}\\
\frac{aH}{k} &{\rm i}\\
\end{array}
\right)
\left(
\begin{array}{c}
a({\bf k})\\
a^{\dag}(-{\bf k})\\
\end{array}
\right).\lb{bog}
\ee
The second piece of the Hamiltonian (\ref{H}), which is proportional
to $a'/a$, is responsible for a mixing 
between creation and annihilation operators. In the Heisenberg
representation it corresponds to  
a Bogolubov transformation; physically it means that particles are produced 
in pairs with opposite momenta. For reasons that will become clear later, this 
phenomenon is called squeezing in the Schr\"odinger picture; the corresponding
squeezing parameter $r_k$ turns out to be given by the expression
\eqref{rk} above. 
>From (\ref{bog}) one can see that mixing of creation and annihilation 
operators is efficient when the 
off-diagonal terms dominate, in other words, on super-Hubble scales
when ${aH}/{k}\gg 1$.  

Using (\ref{yp}) and (\ref{bog}), one obtains after a little algebra,
\be
y({\bf k},\eta) \equiv f_k(\eta)~ a_{\bf k} +
f_k^*(\eta)~a^{\dag}_{-{\bf k}}\ ,\lb{yf} 
\ee
where $a_{\bf k}\equiv a({\bf k},\eta_0)$, and the field modes $f_k$ obey 
Equation (\ref{eq}) and satisfy $f_k(\eta_0)= {1}/\sqrt{2k}$.
At the initial time $\eta_0$, the field modes are deep inside the Hubble radius. 
Equation \eqref{yf} can be written in the suggestive way
\be
y({\bf k},\eta) = \sqrt{2k}~f_{k1}(\eta)~ y_{\bf k} -
\sqrt{\frac{2}{k}}~f_{k2}(\eta) 
~p_{\bf k}\ ,\label{yk}
\ee
where $y_{\bf k}\equiv y({\bf k},\eta_0)$ and $p_{\bf k}\equiv p({\bf
  k},\eta_0)$,  
$f_{k1}= \Re f_k$, $f_{k2}= \Im f_k$. We have in an analogous way momentum 
modes $g_k(\eta)$, with $g_k(\eta_0)=\sqrt{k/2}$,
\be
p({\bf k}) = \sqrt{\frac{2}{k}}~g_{k1}(\eta)~p_{\bf k} + 
\sqrt{2k}~g_{k2}(\eta) ~ y_{\bf k}~.\label{pk}
\ee 
We shall now address the first step in understanding why and to which
extent these quantum field modes appear classically.    

%%%%%%%%%%%%%%%%%%%%%%%%%%%%%%%%%%%%%%%%%%%%%%%%%%%%%%%%%%%%%%%%%%%%%
\section{Quantum-to-classical transition: The pragmatic view}

In the last section we have described the evolution of the quantum modes
in the Heisenberg representation, in which operators evolve 
in time and quantum states do not. 
While the quantum-to-classical transition is in general formulated in the 
Schr\"odinger picture, for the inflationary perturbations the Heisenberg 
picture provides deep insight, too. 

To see this, let us assume that there is a limit in which $f_{k2}$ and $g_{k1}$ 
(or $f_{k1}$ and $g_{k2}$) vanish. Then it is clear from (\ref{yk}) that the 
non-commutativity of the operators $y_{\bf k}$ and $p_{\bf k}$ is no
longer relevant.  
What is the physical meaning of such a limit? Let us consider a
classical stochastic  
system where the dynamics is still described by equations of the form
 (\ref{yk}), but with now $y({\bf k},\eta_0)$ and 
$p({\bf k},\eta_0)$ representing random initial values (c-numbers). 
If $f_{k2}$ and $g_{k1}$ vanish, we get 
\be
\lb{pcl}
p({\bf k},\eta) \equiv p_{\rm cl}(y({\bf k},\eta)) =
\frac{g_{k2}}{f_{k1}}~y({\bf k},\eta)~. 
\ee
This is true for the quantum system (in the operator sense) and for
the classical stochastic system (in the c-number sense).  
Therefore, for a given realization of the perturbation $y({\bf k},\eta)$, the 
corresponding momentum $p_{\rm cl}({\bf k},\eta)$ is fixed and equal to the 
classical momentum corresponding to this value $y({\bf k},\eta)$. 
Then the quantum system is effectively equivalent to the classical 
random system, which is an ensemble of classical trajectories with a 
certain probability associated to each of them \cite{PS96}. 

This is, in fact, what happens for the primordial fluctuations. The
field modes obey (\ref{eq}), and this equation has, on super-Hubble
scales, solutions that become dominant and solutions that become 
negligible (so-called ``growing'' and ``decaying'' modes). 
Eventually the decaying mode can be neglected and one in 
left with the growing mode. It turns out that $f_{k2}$ and $g_{k1}$ are 
decaying modes, and one is left with \eqref{pcl}.

 From the Heisenberg representation it follows that the
operational equivalence with the classical stochastic system does  
not depend on the initial state; this was indeed shown explicitly
for a wide class of initial states (and extended to some gauge-invariant 
quantities) \cite{LPS97}. 
  
We now look at the problem in the Schr\"odinger representation where the 
state evolves in time, while the operators are fixed. 
The initial quantum state of the perturbations is the vacuum state 
$|0,\eta_0\rangle$ satisfying
\be
a_{\bf k}|0,\eta_0\rangle = 0~~~~~~~~~~~~~~\forall {\bf k}~.
\ee
At later times, due to the creation of particles, the time-evolved state 
is annihilated by a more complicated operator,
\be
\Bigl \lbrace y_{\bf k} + {\rm i} \gamma_k^{-1}(\eta) p_{\bf k}
\Bigr \rbrace |0,\eta\rangle = 0~.\label{qst}
\ee
The corresponding (Gaussian) wave function reads
\bea
& & \Psi [ y_{\bf k},y^*_{\bf k},\eta] = \frac{1}{\sqrt{\pi |f_k|^2}} \exp 
\left(-\frac{|y_{\bf k}|^2}{2 |f_k|^2} \lbrace 1-{\rm i}2F(k)\rbrace
\right)\nonumber\\ & & \;
\equiv \left(\frac{2\Omega_{\rm R}(\eta)}{\pi}\right)^{1/4}
\exp\left(-[\Omega_{\rm R}(\eta)+\I\Omega_{\rm I}(\eta)]\vert y_{\bf
    k}\vert^2\right)\ . 
\label{Psi} 
\eea
In (\ref{qst},\ref{Psi}), we have 
\bea
\gamma_k &=& \frac{1}{2|f_k|^2} [ 1 -2{\rm i} F(k)]~,\nonumber\\
F(k) &=& \Im f_k^* g_k~ =  f_{k1} g_{k2} - f_{k2} g_{k1}~.\lb{Fk}
\eea
At the initial time $\eta=\eta_0$, $\gamma_k(\eta_0)=k$, and hence $F(k)=0$;
in other words, we have a minimum uncertainty wave function. This is no 
longer so later, as $|F(k)|$ becomes very large; the probabilities, 
however, remain Gaussian. 
Another way to exhibit the physical meaning of our state is to consider 
the Wigner function, $W$, which can be considered as a kind of
quasi-probability density  
in phase space. For Gaussian wave functions, $W$ has the property to 
be positive definite. For the wave function (\ref{Psi}) one obtains
\be
W  =  \stackrel{|r_k| \rightarrow \infty}{\longrightarrow}  
|\Psi|^2~\delta^{(2)}~\left( p_{\bf k}- p_{\rm cl}(y_{\bf k})
\right)~.\label{wig} 
\ee
The dynamics of the fluctuations leads to the large-squeezing limit
$|r_k| \rightarrow \infty$. 
One gets a highly elongated ellipse whose large axis is oriented along the 
line $p_{\bf k}=p_{\rm cl}(y_{\bf k})$ and whose width becomes
negligible. This is a direct 
vizualisation of the classical stochastic behaviour of our system: the variable 
$y_{\bf k}$ can take any value with corresponding probability $|\Psi|^2$, while 
$p_{\bf k}$ takes the corresponding value $p_{\bf k}=p_{\rm cl}(y_{\bf k})$. 
Instead of being essentially located in phase space around one physical 
trajectory, as for coherent states, the system behaves as if it followed an 
infinite number of classical trajectories with a definite probability to be on 
each of them. 
Interestingly, an analogous situation happens for a free non-relativistic 
particle \cite{KP} possessing an initial Gaussian minimal uncertainty
wavefunction.  
As is well known, $F\propto t$ and becomes very large. At very late times, the 
position does no longer depend on the initial position, 
\be
x(t) \simeq \frac{p_0}{m} t~.\lb{xt}
\ee 
We get an equivalence with an ensemble of classical particles obeying 
(\ref{xt}), where $p_0$ is a random variable with probability 
$P(p_0)=|\Psi|^2(p_0)$. This illustrates the kind of classicality 
we are dealing with. Moreover, when \eqref{xt} holds, position
operators at different times approximately commute (which, in
quantum-optical language, corresponds to a quantum-nondemolition
situation).

Using the canonical commutation relations, the quantum coherence between the 
growing and decaying mode can be expressed as
\be
f_{k1}g_{k1} + f_{k2}g_{k2} = \frac{1}{2}~.
\ee
Clearly, when $f_{k2},g_{k1}$ are unobservable, this coherence becomes 
unobservable as well. This is the case when the decaying mode is so small that we 
have no access to it in observations. For the ratio of the growing to the 
decaying mode one has 
\be
\frac{f_{k2}}{f_{k1}}\propto e^{-2|r_k|}~,
\ee
which is why a large squeezing parameter $r_k$ in the Schr\"odinger
picture implies a vanishing 
decaying mode in the Heisenberg representation. 
The width of the Wigner function is given by
\be
\langle ( p_{\bf k}- p_{\rm cl}(y_{\bf k}))^2 \rangle = g_{k1}^2~,
\ee
which becomes unobservable like the decaying mode. A further
consequence is that the typical phase-space volume occupied by the 
system becomes negligible, too.
 
Let us take the concrete and important example of a perturbation on de~Sitter 
space $a\propto {\rm e}^{Ht}$, with $H$ being constant. The exact solution
of \eqref{eq} with the correct initial condition (ground state for
initial sub-Hubble modes) then reads up to an unimportant {\em constant} 
phase factor
\bea
f_k &=& \frac{-{\rm i}}{\sqrt{2k}}~{\rm e}^{-{\rm i}k\eta} 
\bigl (1-\frac{\rm i}{k\eta}\bigr
)\ ,\\
g_k &=& -{\rm i}~\sqrt{\frac{k}{2}}   {\rm e}^{-{\rm i}k\eta}\ , \quad\eta\equiv
-\frac{1}{aH}<0~.\label{dS}  
\eea
Modes initially inside the Hubble radius become much larger than the 
Hubble radius during inflation solely as a result of their 
dynamics to satisfy $k\eta\ll 1$: here we have the limit mentioned above! 
This can be shown also to correspond to the large-squeezing limit.  
Actually, this is a particular case of the general situation when an
equation like (\ref{eq}) has a  
growing-mode solution and a decaying-mode solution. Here the decaying mode 
becomes vanishingly small; when it is neglected we are in the limit of a 
random stochastic process. Perturbations are then given by 
\be
\delta\phi({\bf k},\eta) = \frac{H}{\sqrt{2k^3}}~ e_{\bf k}~.\lb{delSH} 
\ee
We have set here $\sqrt{2k}~y_{\bf k} = e_{\bf k}$, which assumes the role of 
a classical Gaussian random variable with unit variance. From (\ref{delSH}) we 
see that the perturbations tend to a constant value (they become ``frozen''). 
One should realize that the true reason for the quantum-to-classical transition 
in the sense discussed here is that the decaying mode becomes vanishingly small. 
Primordial gravitational waves follow exactly the behaviour (\ref{delSH}) (up 
to some factor) \cite{S79}, but after re-entering the Hubble radius they will start 
oscillating. They retain their classical appearance because the decaying mode 
(which oscillates as well by then!) is negligible \cite{PS96}. 

%%%%%%%%%%%%%%%%%%%%%%%%%%%%%%%%%%%%%%%%%%%%%%%%%%%%%%%%%%%%%%%%%%
  
\section{Observational predictions}
\par\noindent

The perturbations produced during inflation have remarkable
properties which can be  
confronted with observations. This confrontation makes essential use
of the effective classical behaviour discussed in the last section.

Primordial inflaton fluctuations generate a primordial  
Newtonian potential and the corresponding energy-density fluctuations
$\delta \rho$.  
A central quantity is the power spectrum, $P(k)$, of the quantity 
$\delta\equiv {\delta\rho}/{\rho}$, 
\be
\langle \delta({\bf k})~\delta^*({\bf k}') \rangle = P(k)~\delta^{(3)}
({\bf k}-{\bf k}')~. 
\ee 
When the statistical properties are isotropic, the power spectrum
depends only on  
$k\equiv |\bf k|$. It can be shown that the power spectrum is the
Fourier transform  
of the correlation function (in space), and it can be defined for any
quantity. Deep  
in the matter-dominated stage, $P(k)$ has the following expression on
``super-horizon'' scales in slow-roll single-field inflation,
\be
P(k) = \frac{1024}{75} \pi^3 G^3 \left(
  \frac{V^3}{V'^2}\right)_{t_k}~(aH)^{-4}~k, 
\ee  
where $V'$ is the derivative of the inflaton potential with respect to
the inflaton $\phi$,  
and the fraction has to be evaluated at the Hubble-radius crossing
time $k=a(t_k)H(t_k)$  
\emph{during} inflation. Because of the quasi-exponential inflationary
expansion, it depends very weakly on $k$. Neglecting this dependence, 
we get 
\be
P(k)\propto k \lb{HZ}~,
\ee
which is the scale-invariant ``Harrison--Zeldovich'' spectrum
that plays a crucial role in these investigations. This spectrum is called 
scale-invariant for the following reason: if we compute the
r.m.s. relative mass  
fluctuations $\langle \left({\delta M}/{M}\right)^2 \rangle$ at the time $t_k$
when a scale eventually re-enters the Hubble radius, the same value is
obtained for all scales. 

Using the expansion \eqref{yf} and the commutation relations \eqref{cca}, it is 
straightforward to show that 
\be
\lb{deltaphisquared}
\langle \delta\phi^2 \rangle = \frac{1}{2\pi^2} \int_0^{\infty} dk
~k^2 ~|\delta\phi_{k}(\eta)|^2~, 
\ee
with $f_{k}(\eta)= a ~\delta\phi_{k}(\eta)$. This means that the power
spectrum of $\delta\phi$ is  
just given by $|\delta\phi_{k}(\eta)|^2$. 
However, the average on the left is a quantum average; it is only by
virtue of the  
quantum-to-classical transition mentioned above that we can consider  
$|\delta\phi_{k}(\eta)|^2$ as the power spectrum of a classical random
variable, whose time  
evolution is consistent with probabilities conserved along classical
trajectories. In the  
opposite case this would be impossible due to quantum interferences.  
We note also the result in the limit \eqref{delSH}, which gives
\be 
\frac{d \langle \delta\phi^2 \rangle}{d\ln k}=
\left(\frac{H}{2\pi}\right)^2 ~,
\ee 
where the derivative is with respect to some cut-off value. 

Primordial fluctuations leave their imprint on the CMB and this
provides the best  
constraint on their properties and on the inflationary models in which
they were  
presumably produced. While the CMB is remarkably homogeneous with a black body 
spectrum, perturbations induce very tiny inhomegeneities of the order
$10^{-5}$.   
In this regime, linear perturbation theory is very accurate so that precise 
predictions can be made. The measurement of the temperature
anisotropies angular power spectrum, the $C_{\ell}$'s, 
\be
C_{\ell}=\langle |a_{lm}|^2 \rangle\ , \quad\frac{\Delta
  T}{T}(\vartheta,\varphi)= 
           \sum_{l,m} ~a_{lm} ~Y_{lm}~,
\ee
(which are in the isotropic case independent of $m$) will culminate
with the Planck satellite (ESA).  
The exquisite data we have thus far, in particular those collected by the WMAP
collaboration (NASA), show  
excellent agreement with a flat universe and adiabatic
perturbations \cite{Wmap5,DBHP}. Such perturbations  
respect the equation of state of the background; for the baryon--photon 
plasma this is  
when $\frac{\delta T}{T}=\frac{1}{3}\frac{\delta n_B}{n_B}$, where $n$
is the baryon number  
density. This is a natural outcome of single-field inflation. 

Before decoupling, the  
baryon--photon plasma is tightly coupled and its density 
oscillates on scales inside the Hubble radius, yielding oscillations
similar to pressure  
waves. These are often called acoustic oscillations. The
location of the first (Doppler) peak  
gives roughly the angular scale of the Hubble radius at decoupling and
is consistent
with a flat universe. The pattern of the angular power spectrum is in
agreement with  
primordial adiabatic fluctuations. After decoupling, the baryons
retain the primordially  
induced acoustic ``Sakharov'' oscillations, the baryonic acoustic
oscillations (BAO);  
these were detected in the galaxy power spectrum and are presently
used in order to constrain dark energy models.

To parametrize the departure from scale invariance, one introduces the
spectral index $n$ with  
$P(k)\propto k^n$. Latest CMB data constrain $n$ to be very close, but
slightly lower than one \cite{Wmap5}. 
Finally we see no clear evidence for non-Gaussianity in the
statistics of the perturbations.  
All these data are in surprisingly good agreement with the simplest
single-field slow-roll inflationary models (see e.g. \cite{KKMR08}). 

Let us return in more detail to the acoustic oscillations. They arise
because of the  
standing-wave behaviour of the perturbations inside the Hubble
radius. There are always  
two modes that are solutions to the equations and they will both
oscillate. One of the  
modes matches the growing (dominant) mode, and the other the decaying
(subdominant) mode.  
For modes sufficiently long outside the Hubble radius, the decaying
mode disappears and  
the growing mode will match the corresponding oscillating mode inside
the Hubble radius.   
At decoupling, each mode has a given oscillation phase, and this gives
rise to the acoustic oscillations seen in the $C_{\ell}$'s. 
If we had a way  
to generate classical perturbations that would evolve outside the
Hubble radius for very  
long, just the same would be true. If these perturbations had random
initial conditions,  
obeying the same statistics as our initially quantum fluctuations,
both systems would be  
indistinguishable. Hence the presence of acoustic oscillations is in
no way connected to  
the quantum nature of the perturbations but rather to their primordial
origin. But the  
quantum-to-classical transition can only take place in a system where
the decaying mode  
is negligible enough so that acoustic oscillations {\em do} arise. It is
interesting that a similar  
standing-wave behaviour is present in the primordial stochastic
gravitational waves  
background produced during inflation. Unfortunately, to detect it in a
direct detection  
experiment today would require a resolution in frequency of about
$10^{-18}$ Hz, \cite{PS96} clearly 
beyond present or foreseeable capabilities. 
The same property yields also small superimposed oscillations in the
power spectra of the CMB temperature anisotropy and polarization. This is 
similar to the acoustic oscillations but with a period approximately
twice as small
(solely due to the difference between the light velocity and the sound velocity in 
the baryon--photon plasma at the recombination time) \cite{PS96}. Their observation 
is very difficult but not hopeless if the parameter characterizing the tensor-to-scalar 
ratio in the CMB temperature anisotropy is not too small, see \cite{LPPS00}
for detailed estimates of the CMB polarization B-mode produced by primordial 
gravitational waves only. 

We finally mention that
calculations done for the creation of matter by parametric resonance after 
inflation use the description of perturbations in terms of classical
stochastic fields.   
All the predictions mentioned above and which were confirmed by
observations are done  
in the closed-system approach, that is, by taking the perturbations as
an isolated system. 
Similar results were obtained in various disguise by several authors 
\cite{GP85,L85,AFJP94} and even extended beyond the linear regime 
\cite{LS08}.
In this approach the system becomes indistinguishable, in an
operational sense, from a classical stochastic system solely by 
virtue of its peculiar inflationary dynamics.

 From a purely pragmatic point of view, the closed-system approach is
sufficient. In astrophysical observations one measures certain
classical correlation functions for which the above line of thought 
shows that they are indistinguishable from the fundamental quantum 
expectation values. Still, in the next section we shall go beyond the
closed-system approach by taking into account the interaction of the
modes with other, ``environmental'', degrees of freedom. 
This has several reasons. 
First, the environment-induced decoherence process is generally invoked 
in order to explain the appearance of classical behaviour in quantum 
theory \cite{deco}. 
Second, since an environment is expected to be present anyway, it is 
important to consider whether it does not spoil the successful predictions 
from the closed-system approach. It should, in particular, not erase the 
acoustic oscillations. Moreover, invoking large non-linear effects might 
irremediably modify the CMB angular power spectrum and induce large 
non-Gaussianity.
Finally, there is the question about the entropy of the perturbations 
which by definition cannot be addressed inside the closed-system approach.

We shall see that these questions and problems can be successfully dealt 
with without spoiling the successful predictions of the closed-system 
approach including the quantum-to-classical transition in the pragmatic 
approach adopted in this section.  

%%%%%%%%%%%%%%%%%%%%%%%%%%%%%%%%%%%%%%%%%%%%%%%%%%%%%%%%%%%%%%%%%%%%%%%

\section{Quantum-to-classical transition: decoherence}

\subsection{Decoherence and pointer basis}

In the last section we have described the primordial fluctuations in
cosmology by a collection of independent quantum states labelled by
the wave number $k$. Since no interaction between different $k$ or
between the fluctuations and other fields have been considered, we
deal with a pure quantum state for each $k$. The initial condition for
each quantum state is the harmonic-oscillator ground state with
respect to $k$. During inflation, modes with wavelengths larger
than the Hubble scale $H^{-1}$ assume a squeezed Gaussian state. We
focussed attention on the modes far outside the Hubble scale, which
experience an enormous squeezing. For these
highly-squeezed modes, which are the ones relevant for cosmological
observations, all expectation values containing the
field-amplitudes or their momenta are indistinguishable from classical
stochastic averages \cite{PS96}. It is this approximate coincidence
between quantum and classical expectation
which is the basis of the pragmatic approach to the
quantum-to-classical transition discussed above for the primordial
fluctuations.  

One can, however, adopt a more fundamental point of view.
It is far from realistic to
assume that a primordial fluctuation with wave number $k$ is exactly
isolated. We must take into account its interaction with other degrees
of freedom (called the `environment' for simplicity).
The main reason is the following. As one knows from
standard quantum theory, even a tiny interaction with other degrees of
freedom can become important, in the sense that an entanglement of a
system with its environment can form even without direct disturbance
of the system. If the environmental degrees of freedom are
inaccessible to observations (as they usually are), the ensuing
entanglement with the system leads to {\em decoherence} --
interference terms can no longer be observed at the system itself and 
the system {\em appears} classical \cite{deco}.
 This is the fundamental origin of the
quantum-to-classical transition. The phenomenon of decoherence is by
now theoretically 
well understood and has been experimentally tested with high precision
\cite{deco,zurek,schlosshauer}.
Decoherence leads to an {\em apparent
ensemble} of wave packets for the observable with respect to which
the interferences vanish. A paradigmatic example is the localization
of a quantum particle due to scattering with photons, air molecules,
or other particles \cite{deco,schlosshauer,JZ}. There the position
basis of the particle is the approximate basis distinguished by the
scattering process. The basis distinguished by the environment is 
generally called the {\em pointer basis}; 
the corresponding observable is called pointer observable.
 Interferences between different members of the
pointer basis are suppressed by the decohering influence of the
environment. 

One would expect, therefore, that decoherence is of crucial importance
for the primordial fluctuations, too. This expectation is, moreover,
supported by the fact that the system by itself evolves into a highly
squeezed state in which squeezing is in the field momentum and
broadening is in the field amplitude (corresponding to the position
variable in quantum mechanics):
 one knows from quantum theory that highly squeezed states are
 extremely sensitive to any environment \cite{deco}. This is the reason
 why they are so difficult to generate in the laboratory -- it is very
 hard to isolate them from any environment. In view of their huge
 squeezing, this argument should apply to the cosmological
 fluctuations {\em a fortiori}.

But could it be imaginable that the cosmological fluctuations, in
contrast to a typical quantum-mechanical situation, are indeed
strictly isolated? The answer is definitely no. 

Firstly, in any fundamental theory (such as
string theory) there is an abundance of different fields with different
interactions. Among them it will not be difficult to find appropriate
candidates for environmental fields generating decoherence for the
primordial fluctuations.

Secondly, even if one assumes to have no such fields,
there are two processes which cannot be neglected. The first one is
the interaction between modes with different $k$; recall that the full
theory is non-linear and that, therefore, the various modes cannot be
treated independently of each other. Such non-linear interactions
concern both the interaction with the modes of the inflaton and the
perturbations of the metric (containing, in particular, gravitational
waves).

The second process is the entanglement
of the modes' quantum state between different {\em spatial regions}:
even if the modes are independent in $k$-space, the Gaussian wave
functions for the amplitudes in real space are highly correlated over
spacelike regions (as in the Einstein--Podolsky--Rosen
situation). This leads, in 
particular, to an entanglement between
the regions inside and outside the Hubble radius. Famous
non-cosmological examples are the Hawking and the Unruh effects, where
the thermal appearance of the corresponding radiation can be
understood from the entanglement between inside and outside the event
horizons and the tracing out of the correlations into the horizon
\cite{israel}. Even for spacelike surfaces which stay outside the
horizon, the thermal nature of Hawking and Unruh radiation can be
understood from the entanglement with other fields, leading to
decoherence \cite{kiefer01}.

The process of decoherence is, moreover, needed to justify the results
from the isolated (closed) system in the first place. Even if the classical and
quantum expectation values are indistinguishable, the presence of a
pure state means that one has a quantum superposition of all possible
field amplitudes, {\em not} an ensemble of stochastically distributed
classical values. This situation is similar to Schr\"odinger's cat.
In the pragmatic point of view of Sec.~III, the
approximate coincidence of the expectation values suffices. 
Such a coincidence
 is, however, not sufficient for a realistic interpretation. Only
decoherence can eventually justify the pragmatic point of view in that it
leads to an apparent ensemble of wave packets for the system variables
itself (which, in our case, are the field amplitudes).
The insufficience of
approximately equal classical and quantum expectation values for a
fundamental interpretation has
recently been clearly emphasized in a different context (the quantum
mechanics of classically chaotic systems) by Schlosshauer
\cite{hyperion}. In the presence of a pure state one can always find
an observable for 
which no classical counterpart exists, that is, for which the
comparison of quantum and classical expectation values is
meaningless. 

The quantum-to-classical transition happens for the highly-squeezed
modes whose wavelengths exceed the Hubble scale. It is for these modes
where environmental decoherence is most efficient \cite{KPS1}. How can
this happen? Would one not expect that no causal interaction can
occur on scales larger than the Hubble scale? This is
true only for a direct disturbance of the system. But the crucial
point is that quantum entanglement can form without direct disturbance. And
this is all one needs for decoherence! In the context of the quantum
measurement process, the sole formation of entanglement is referred 
to as an `ideal measurement' or a `quantum non-demolition measurement': 
the system remains undisturbed, but the environment is affected through the
formation of entanglement. The general mechanism is as follows \cite{deco}.

Consider a quantum system which
is initially in the state $|n\rangle$ and a
`measurement device' (here: the environment) which is in
some initial state $|\Phi_0\rangle$. (We assume that $|n\rangle$
belongs to a set of
eigenstates of a system observable.)
The evolution according to the Schr\"odinger equation
is in the special case of an `ideal measurement' given by
\be |n\rangle|\Phi_0\rangle \stackrel{t}{\longrightarrow}
     \exp\left(-{\rm i} H_{\rm int}t\right)|n\rangle|\Phi_0\rangle
     =|n\rangle|\Phi_n(t)\rangle\ ,  \label{ideal} \ee
where $H_{\rm int}$ denotes the interaction Hamiltonian (assumed here
to dominate over the free Hamiltonians) which
correlates the system state with its environment without changing the
system state.

In the general case, the quantum system can be in a superposition of
different eigenstates of the system observable. Then, due to the
linearity of the time evolution, an initial product state with
$|\Phi_0\rangle$ develops into an entangled state of system plus apparatus,
\be \left(\sum_n c_n|n\rangle\right)|\Phi_0\rangle
    \stackrel{t}\longrightarrow\sum_n c_n|n\rangle
    |\Phi_n(t)\rangle\ . \label{measurement}
\ee
But this is a highly non-classical state! Since the
environmental states $\{\vert\Phi_n\rangle\}$
are not accessible, they have to be traced out
from the full quantum state. One thereby arrives at the reduced
density matrix $\rho_{\rm S}$
which contains all the information that is available at
the system itself. Since the environmental states
$\{\vert\Phi_n\rangle\}$
can be assumed as being approximately orthogonal (otherwise they would not be
able to serve as a `measurement device'), the reduced density matrix
is of the form
\be
\rho_{\rm S} \approx \sum_n |c_n|^2 |n\rangle\langle n|,
    \label{deco}
\ee
that is, it assumes the form of an {\em approximate ensemble} for the
various system states $\vert n\rangle$, each of which occurs with
probability $|c_n|^2$.

In our case, the cosmological fluctuations represent the system to be
decohered. The environmental states $\{\vert\Phi_n\rangle\}$ can be
other fields or inaccessible parts of the fluctuations themselves
(see below). The system states $\vert n\rangle$ are given by the
field-amplitude states $\vert y_{\mathbf k}\rangle$. The interaction with the environment
can, in the ideal-measurement case,
be described by the multiplication of an initial density matrix
$\rho_0(y,y')$ with a Gaussian factor in $y-y'$ (omitting here and in
the following the index ${\mathbf k}$ in $y_{\mathbf k}$),
 \be
\lb{rhoxi} \rho_0(y,y')\longrightarrow
\rho_{\xi}(y,y')=\rho_0(y,y')
\exp\left(-\frac{\xi}{2}(y-y')^2\right). \ee
Here, the parameter $\xi$ encodes the details of the interaction
between the modes and their environment. Given a specific model with a
specific interaction, $\xi$ can be calculated. The special decoherence
process \eqref{deco} is typical for the description of localization in
quantum mechanics \cite{JZ,deco,schlosshauer}.

One recognizes from \eqref{rhoxi} that interferences between different
values of the field amplitude $y$ have been suppressed by interaction
with the environment. This is decoherence. So far we have just
assumed without derivation
that $\vert y\rangle$ is the pointer basis, that is, the relevant
robust system basis which is distinguished by the environment. This
must, of course, be justified. A detailed derivation for the
field-amplitude basis to be the pointer basis has been presented in
\cite{KPS1} and \cite{KLPS2}. We review here the main arguments and
refer the reader to these references for more details.

According to the classical equations, for modes with very large
wavelength one has $y\propto a$, that is, the physical fluctuations
$\delta\phi$ are approximately constant (`frozen'). In the Heisenberg
picture of the quantum theory, this means that the operator
$\widehat{\delta\phi}$ approximately commutes with the
Hamiltonian. Now comes the crucial 
point. Additional (environmental) fields
coupling with the cosmological fluctuations are expected to couple
field amplitudes, not canonical momenta of field amplitudes; that
is, the coupling is expected to involve $\widehat{\delta\phi}$, {\em not}
its momentum. Consequently, the fluctuations $\widehat{\delta\phi}$
commute with the {\em whole} Hamiltonian of system plus
environment. Such a variable is a pointer observable par excellence
\cite{deco,zurek,schlosshauer}. It is stable (robust) in time because
of this commutativity which holds for the wavelengths much bigger than
the Hubble scale. The phenomenological expectation \eqref{rhoxi} is
thus fully justified. One must keep in mind, though, that
$\widehat{\delta\phi}$ is only an approximate pointer observable: although
the non-diagonal terms in \eqref{rhoxi} become exponentially
suppressed, they never vanish exactly, as would be the case if the
$\widehat{\delta\phi}$ 
were the exact pointer observable. In fact, the
reduced density matrix can be decomposed into narrow Gaussians in
$\delta\phi$-space. The 
whole situation is in strong analogy to the
localization of a massive particle by scattering with the environment
\cite{deco,schlosshauer,JZ}.

The approximate commutativity of $\widehat{\delta\phi}$ with the full
Hamiltonian means in particular that the kinetic term, that is, the
$p^2$-term, of the system becomes irrelevant in the large-squeezing
limit. If this term were relevant (as it is for modes with smaller
wavelength), the pointer basis would not be the field-amplitude basis,
but the coherent-state basis \cite{KLPS2}. But this is not the case
here. The coherent-state basis is, in particular, unstable under the
time evolution.

So far we have restricted our attention to a special initial state:
the vacuum state. This is, however, not necessary. In \cite{KLPS2} we
have presented a 
formalism that is general enough
to encompass a wide range of initial states and interactions. A
central role in this 
formalism is played by a master equation for the reduced density
matrix, which is of the Lindblad form. More concretely, the density
matrix is assumed to satisfy \cite{deco}
\be
\lb{Lindblad}
\frac{\D\hat{\rho}}{\D t}=-\I[\hat{H},\hat{\rho}]
+\hat{L}\hat{\rho}\hat{L}^{\dagger}-\frac12\hat{L}^{\dagger}\hat{L}\hat{\rho}
-\frac12\hat{\rho}\hat{L}^{\dagger}\hat{L}\ ,
\ee
where $\hat{L}$ is the Lindblad operator. Most of the particular
models discussed in the literature lead to a master equation of this
form. It is thus of interest to study this equation as general as
possible. We have assumed that the Lindblad operator
is linear in our variables $p$ and $y$, but kept it general otherwise.
 The Hamiltonian $\hat{H}$ is given by the expression \eqref{H}.

The results of our discussion in \cite{KLPS2} can be summarized as
follows. It turns out that the behaviour of the master equation is qualitatively
different for modes outside the Hubble radius (as is
the case here) and the modes inside. The decoherence time $t_{\mathrm d}$
for the modes with wavelengths much bigger than the Hubble radius is
during inflation of the order
\be
\lb{td}
t_{\mathrm d} \sim H_{\rm I}^{-1}\ln\frac{H_{\rm I}^{-1}}{t_0},
\ee
where $H_{\rm I}$ is the (approximately constant) Hubble parameter
during inflation, and $t_0$ is a typical time characteristic
for the details of the interaction. We emphasize that \eqref{td} is approximately
independent of these
details. It is basically given by the Hubble time,
with the details only entering logarithmically. The time $t_{\rm d}$
also gives the timescale for the Wigner function to become
positive. The reduced density matrix can then be decomposed into an
apparent ensemble of narrow Gaussians for the values of the field
amplitude, cf. \cite{DK} for a general discussion. For the
large-wavelength modes in the radiation-dominated phase one obtains instead
\be
\lb{tdr}
t_{\mathrm d} \sim \frac{H_{\rm I}t_{\rm L}^2}{2},
\ee
where $t_{\rm L}$ depends again on the details of the interaction. One
has now a more sensitive dependence on the interaction. Moreover, for $H_{\rm I}t_{\rm
L}\gg 1$ one has a much longer decoherence time than
during inflation. This means that, depending on the interaction, 
decoherence can be much less efficient than during inflation.

For modes smaller than the Hubble scale, the situation is very
different \cite{KLPS2}. Taking as a representative example a photon
bath as the environment (realized e.g. by the CMB), the decoherence
time is independent of the Hubble parameter and strongly dependent on
the coupling to the bath. Dissipation now becomes the dominant source
of influence, in contrast to the case of the super-Hubble modes for
which only entanglement occurs.

Decoherence is often connected with symmetry breaking
\cite{deco}, see also \cite{Zeh}, section~6.1.
 This is also the case here. The initial de~Sitter-invariant vacuum state for
the fluctuations is highly symmetric. But the observed classical
fluctuations are certainly non-symmetric. This can easily be
understood and does not require new physics (as e.g. demanded in
\cite{PSS}). The initial vacuum state develops into a squeezed vacuum,
which can be understood as a superposition of different
field-amplitude eigenstates. Decoherence then makes this
indistinguishable from an ensemble of (approximate) field-amplitude
eigenstates, each of which is highly inhomogeneous. The situation
resembles the case of spontaneous symmetry breaking in field theory,
where the symmetric initial state evolves into a superposition of
`false vacua'. After decoherence one is left with an apparent ensemble
of different false vacua, one of which corresponds to our observed
world.

\subsection{Entropy}

In Sec.~III the primordial fluctuations were treated as isolated and
thus described by a pure (squeezed) state. Consequently, they possess
zero entropy: all information is contained in the system itself. But as we
have seen, the primordial fluctuations are an {\em open} quantum
system; they are entangled with their environment. Because of this
entanglement, the fluctuations are described by the reduced density
matrix \eqref{rhoxi}. They thus possess positive entropy because the
information about the correlations with the environment are
unavailable in the system itself. The local entropy is calculated from
the standard von Neumann formula,
\be
\label{entropy}
S=-{\rm tr}(\rho_{\xi}\ln\rho_{\xi})\ ,
\ee
where $\rho_{\xi}$ is given in \eqref{rhoxi}, and 
where we have set $k_{\rm B}=1$. Considering one (real) mode with wave
number $k$, the maximal entropy, $S_{\rm max}$, would be $2r_k$, where $r_k$
is again the squeezing parameter \cite{Prokopec} (we skip again the
index $k$ in the following).  
We have calculated
and discussed the entropy for the fluctuations in \cite{KPS2,KLPS2}. 
To display the result, it is convenient to introduce the
dimensionless
parameter $\chi=\xi/\Omega_{\rm R}$, where $\Omega_{\rm R}$ is the
width of the Gaussian \eqref{Psi}; it controls the strength of decoherence.
 (In the case of pure exponential
inflation one has $\chi=\xi(1+4\sinh^2r)/k$.)
Inserting \eqref{rhoxi} into \eqref{entropy},
one gets the explicit expression \cite{KLPS2}
\be\label{S-exact}
\begin{split}
S&=-\ln\frac{2}{\sqrt{1+\chi}+1}-\frac12\left(\sqrt{1+\chi}-1\right)
                \ln\frac{\sqrt{1+\chi}-1}{\sqrt{1+\chi}+1} \\
 &=\ln\frac12\sqrt{\chi}
                -\sqrt{1+\chi}\ln\frac{\sqrt{1+\chi}-1}{\sqrt{\chi}}\ .
\end{split}
\ee
One recognizes that the entropy vanishes for $\xi\to 0$, as it must
for a pure state. 
In the limit $\chi\gg1$ (large decoherence) one gets
\be
\lb{S-approx}
S=1-\ln2+\frac{\ln\chi}{2}+{\mathcal O}(\chi^{-1/2}) \ .
\ee
This asymptotic value is readily attained. 

As we have emphasized above, modes with wavelength bigger than the
Hubble scale can only experience pure entanglement, not direct
disturbance. In such a case the entropy obeys the bound
\be
\lb{bound}
S < \frac{S_{\rm max}}{2}=r\ .
\ee
The same bound follows from the general discussion of the Lindblad
equation \cite{KLPS2}. It can also be interpreted in the following way
\cite{KLPS2}: in spite of decoherence, some squeezing compared to the
vacuum state (which has $\Omega_{\rm R}=k$) remains. In the language of
the Wigner function it means that the Wigner ellipse is not smeared
out to become a circle, but still exhibits an elongated and a squeezed
part. And this has important consequences for observation! If the
bound \eqref{bound} were violated, there would no longer be any
coherences between the field amplitude and the momentum and,
consequently, no coherences in the coupled baryon--photon plasma
(Sec.~IV). There would then not be any acoustic peaks in the
anisotropy spectrum of the CMB -- in contrast to observation! The
fundamental questions of the quantum-to-classical transition have thus
observational relevance. 

The upper bound $S_{\rm max}/2$ corresponds to the case when the
pointer basis is the exact field-amplitude basis. 
(For $S=S_{\rm max}$, the pointer basis would be the particle-number
basis.) As our pointer basis
consists of narrow packets in field amplitudes, the entropy of the
fluctuations approaches the upper bound asymptotically. 

The existence of the bound \eqref{bound} shows, again, how peculiar
the case of fluctuations in an inflationary universe is. According to
a theorem by Page \cite{Page} (see also \cite{hayden}), if a total
quantum system with dimension $mn$ is in a random pure state, the
average entropy of a subsystem of dimension $m\leq n$ is almost
maximal. But this is not the case for our system: the situation for the
fluctuations during inflation {\em is} very special, and their entropy
cannot exceed half of the maximal entropy, which leaves enough
information for the formation of the acoustic peaks. 

Our results for the entropy in \cite{KPS2} and \cite{KLPS2} also yield
the following simple formula for the entropy production during
inflation:
\be
\lb{production}
\dot{S}\approx \dot{r}\approx H_{\rm I}\ .
\ee
For chaotic systems, the entropy production rate is proportional to
the Lyapunov parameter. This would correspond in our case to the
Hubble parameter $H_{\rm I}$. However, our system is not chaotic, but
only classically unstable, so the analogy is not complete. 

Using \eqref{td}, one can find the amount of entropy produced after
the decoherence time $t_{\rm d}$,
\be
S \sim H_{\rm I}t_{\rm d} \sim \ln\frac{H_{\rm I}^{-1}}{t_0}\ .
\ee
In the radiation-dominated phase following inflation, a relation
similar to \eqref{production} holds, with  $H_{\rm I}$ replaced by the
Hubble parameter $H\propto t^{-1}$. The entropy thus only increases
logarithmically in time, not linearly as in inflation.

\subsection{Specific models}

So far, we have kept the discussion as general as possible. We have
reviewed the arguments which lead to the result that cosmological
fluctuations appear like a classical ensemble of field
amplitudes. Necessary requirements are the inflationary expansion of
the universe and the focus on modes that are highly squeezed.
An interaction with some environment is needed, but the details of it
are unimportant. Still, it is of interest to discuss specific examples
for such interactions. Our paper \cite{KLPS2} gives an extended list of
references; here we shall restrict ourselves to some recent examples.

The purely spatial entanglement between the modes inside the Hubble
scale and outside the Hubble scale was discussed in \cite{SM}, see
also \cite{Nambu}. It was shown there that this entanglement is, by
itself, sufficient to produce the desired decoherence. This is
analogous to the black-hole case
where the decoherence from the tracing out of the modes behind the
horizon leads to the thermal radiation of
the Hawking effect \cite{israel,kiefer01}. The authors of \cite{SM}
also showed that the entropy scales with the volume inside the Hubble
scale and satisfies an upper bound of $S\approx r$ per mode, which
coincides with the 
upper bound \eqref{bound} discussed above. It is thus not in
conflict with the observed acoustic peaks in the cosmic microwave
background.

Instead of pure spatial entanglement one can consider the entanglement
of our strongly squeezed super-Hubble modes with sub-Hubble modes
(which then play the role of the environment). This was discussed, for
example, in \cite{BHH}. The authors take the short-wavelength modes to
be in their ground states and find that decoherence is not sufficient
during inflation. This happens because vacuum states are
usually ineffective to lead to decoherence \cite{deco}. Our arguments
above and in \cite{KLPS2} can thus only be applied to this model 
if at least some modes are not in their ground states. But such modes
can be found: one can interpret the fluctuations with wavelengths
$\lambda\gtrsim H_{\rm I}^{-1}$ as an appropriate environment; they
assume a role intermediate between ground state and state with large
squeezing. Ideas similar to the ones in \cite{BHH} have been pursued
in \cite{martineau,LN}, and elsewhere, with results that are
consistent with our general discussion above. 
A variant of this system-environment split is presented in
\cite{PR} using a two-field model of inflation. 
There, the system consists of curvature perturbations, and
the environment consists of isocurvature modes. 
Finally, another possible source of sub-Hubble modes being in non-vacuum 
states is the {\em secondary} gravitational wave background (``foreground'' 
in astronomical terminology) emitted by matter after the end of inflation 
\cite{KPS2}. 

%%%%%%%%%%%%%%%%%%%%%%%%%%%%%%%%%%%%%%%%%%%%%%%%%%%%%%%%%%%%%%%%%%%%%
\section{Conclusion}

Inflation is a robust scenario which gives an elegant solution  
to some oustanding problems of Big-Bang cosmology, and its predictions
are in agreement  
with present observations, in particular the accurate CMB anisotropy data. It 
is gratifying that this scenario offers also the possibility to deal with such 
fundamental and subtle questions as to why quantum perturbations
produced in the  
early Universe give rise to classical inhomogeneities today. We
believe that this aspect  
is no less fascinating than its other successful predictions. 

We expect that  models of the quantum-to-classical transition for
the primordial fluctuations will continue to appear in the
literature. But we are convinced that the general mechanism of this
transition presented in this essay will hold true for all scenarios
based on inflation.    

%%%%%%%%%%%%%%%%%%%%%%%%%%%%%%%%%%%%%%%%%%%%%%%%%%%%%%%%%%%%%%%%%%%%

\section*{Acknowledgements}

We are happy to thank our collaborator Alexei Starobinsky for his crucial 
input in obtaining the results presented here and for his comments on
our manuscript.

We kindly acknowledge financial support from The Foundational Questions
Institute (http://fqxi.org) for the visit of D. P. to the University
of Cologne.

%%%%%%%%%%%%%%%%%%%%%%%%%%%%%%%%%%%%%%%%%%%%%%%%%%%%%%%

%%%%%%%%%%%%%%%%%%%%%%%%%%%%%%%%%%%%%%%%%%%%%%%%%%%%%%%%

\end{document}